\documentclass{aa}
\usepackage{bm} %bold math (vectors)
\usepackage{txfonts}
\usepackage{natbib}
\usepackage{nicefrac}
\usepackage{graphicx}
\usepackage{hyperref}            
\usepackage{placeins}
\usepackage{multicol}

\hypersetup{colorlinks = {true},
            linkcolor = [rgb]{0,0.35,0.7},
            citecolor = [rgb]{0,0.35,0.7},
            filecolor = [rgb]{0.61,0,0},
            urlcolor = [rgb]{0.61,0,0},
           }

\graphicspath{ {figures/} } %place figures there

\newcommand*\dd[1]{\,\mathrm{d}#1}

\newcommand{\tensorGR}[1]{\overline{\bm{{#1}}}}
\newcommand{\DP}[2]{\frac{\partial{#1}}{\partial{#2}}}

\newcommand{\bin}{_\mathrm{bin}}
\newcommand{\cav}{_\mathrm{cav}}
\newcommand{\G}{\text{G}}
\newcommand{\Mco}{M_\text{co}}
\newcommand{\Msun}{\text{M}_\odot}
\newcommand{\Rgas}{\mathcal{R}}
\newcommand{\cs}{c_\text{s}}
\newcommand{\ciso}{c_\mathrm{s,iso}}

\newcommand{\OmegaK}{\Omega_\text{K}}

\newcommand{\cv}{c_\text{v}}
\newcommand{\sigmaSB}{\sigma_\text{SB}}
\newcommand{\vel}{\bm{u}}
\authorrunning{Sudarshan et al.}

\begin{document}

\title{How cooling influences circumbinary discs}

\author{Prakruti Sudarshan\inst{\ref{inst1}},
		Anna~B.~T.~Penzlin\inst{\ref{inst1}},
        Alexandros Ziampras\inst{\ref{inst1},\ref{inst2}},
    	Wilhelm Kley\thanks{W.~Kley is deceased and is included as a co-author for his significant contribution and guidance over the course of this project.}\inst{\ref{inst1}} \and
    	Richard~P.~Nelson\inst{\ref{inst2}}}

\institute{
Institut f\"ur Astronomie und Astrophysik, Universität T\"ubingen,
Auf der Morgenstelle 10, D-72076, Germany \label{inst1}
	\and Astronomy Unit, School of Physics and Astronomy, Queen Mary University of London, London E1 4NS, UK \label{inst2}\\
\email{prakruti.sudarshan@student.uni-tuebingen.de, \{anna.penzlin, alexandros.ziampras\}@uni-tuebingen.de}}

\date{}

\abstract
{
Circumbinary disc observations and simulations show large, eccentric inner cavities. Recent work has shown that the shape and size of these cavities depend on the aspect ratio and viscosity of the disc, as well as the binary eccentricity and mass ratio. It has been further shown that, for gaps created by planets, the cooling timescale significantly affects the shape and size of the gap. In this study, we consider the effect of different cooling models on the cavity shape in a circumbinary disc. We compare locally isothermal and radiatively cooled disc models to ones with a parametrised cooling timescale ($\beta$-cooling), implemented in 2D 
numerical simulations for varying binary eccentricities. While the shape of the cavity for radiative and locally isothermal models remains comparable, the inner disc structure changes slightly, leading to a change in the precession rate of the disc. With $\beta$-cooled models, the shape and size of the cavity changes dramatically towards values of $\beta=1$. Based on our findings, we introduce a parametrised $\beta$ model that accounts for the shorter cooling timescale inside the cavity while adequately reproducing the results of the radiative model, and we highlight that accurate treatment of the thermodynamics inside the cavity has a significant impact in modelling circumbinary systems.
}

\keywords{
          Hydrodynamics --
          Methods: numerical --
          Binaries: general --
          Accretion, accretion discs --
          Planets and satellites: resonance --
          Protoplanetary discs
         }

\maketitle
\section{Introduction} \label{sec:intro}

About 50\% of all stars are part of a binary configuration. Binary stars are believed to form together often, such that they have a common circumbinary disc in which planets can form and migrate. The Kepler mission has revealed about a dozen close-in circumbinary planets \citep{2011Kepler16, 2012Kepler38, 2019Kepler47, 2012Kepler34-35, 2015Kepler453, 2013Kepler64, 2014Kepler413, 2020TOI-1338, 2020Socia}. The circumbinary protoplanetary disc (CBD/PPD) is key to understanding the evolution of the binary stars and circumbinary planets during their formation.

The structure of the disc and the interaction between discs and embedded objects is a subject of ongoing research. In the context of binary super-massive black holes (SMBHs), accretion from the disc onto the binary has been studied using various models. The effect of the disc on a binary object was recently studied by \cite{2019Munoz, 2020Munoz} and \citet{2020Tiede}, discovering binary expansion for thick, viscous discs ($\alpha\geq 0.01$, $h=0.1$). \cite{2021Tiede} looked into the accretion dynamics with tracer particles embedded in a Cartesian grid simulation. The effects of the binary eccentricity on the momentum exchange between the binary components was investigated in models by \cite{2021Dorazio}. \cite{2021Dittmann, 2022Dittmann} explored the effect of the disc height and binary mass ratios. A 3D study by \cite{2019Moody} tested how misalignment between the disc and binary changes the accretion rate. In this context of SMBHs, while disc heights are comparable to PPDs, the viscosity is much higher, leading to a faster evolution of the models, but also making the handling of viscous heating in the disc even more relevant.

\cite{2019Moesta} studied the flow inside the inner eccentric cavity of a thick disc ($h=0.2$) and found that the stream pattern inside the inner cavity is likely the product of spiral shocks. 
For PPD-like systems, the recent work by \citet{2017Thun}, \citet{2020Hirsh}, and \citet{2020Ragusa} shows large eccentric inner cavities in circumbinary discs, the properties of which depend on the disc and binary parameters.
The migration of planets in circumbinary discs was studied in \cite{2018Thun} and \citet{2021Penzlin} without self-gravity and by \cite{2017Mutter} with self-gravity.

All of the mentioned studies so far have used vertically integrated, locally isothermal models, where the cooling timescale is negligible compared to the orbital period of the gas, ignoring changes to the thermal profile of the disc.
\cite{2019Kley} introduced a radiatively cooled model for the circumbinary disc with migrating planets and \cite{2020Pierens, 2021Pierens} used a three-dimensional (3D) model with $\beta$- cooling including dust to study the impact of the turbulent disc on in situ planet growth close to the binary.

These studies confirm large eccentric inner cavities that are also derived analytically in \cite{2020Munoz_theo}. One caveat to simulating the disc, however, is the long time needed to fully evolve this inner cavity to a convergent state. As a result, simulations that include many advanced physical models can be unfeasible due to runtime constraints. The locally isothermal prescription can provide significant speed-up, at the cost of excluding the thermal evolution of the disc and cavity, which can be critical to their equilibrium state. As a result, the effects of a thermal model on the disc have not been compared in detail.

For single-star protoplanetary discs, recent work by \cite{2020Miranda_I, 2020Miranda_II} and \cite{2020Alex_II} has shown that the cooling timescale can change the gap opening capabilities of embedded planets and the density slope at their gap edge, due to radiative cooling interfering with the angular momentum transport efficiency of spiral arms launched by such planets. Such effects could become relevant for the circumbinary cavity edge and therefore for the inner disc shape.

In this paper, we investigate how the locally isothermal model compares to the simple approach of a parametrised, constant cooling timescale and a more complex radiatively cooled model in the context of PPDs. We also construct simplistic models with adaptive parametrised cooling that aim to reach a compromise between computational speed and physical accuracy.

In Sect.~\ref{sec:model}, the model setup is explained and the radiative effects considered are described. The results are shown in Sect.~\ref{sec:results}, where all models are compared and contrasted throughout our parameter space. We discuss our findings as well as possible caveats and improvements in Sect.~\ref{sec:discussion}. Finally, our results are summarised in Sect.~\ref{sec:conclusions}.

\section{Model setup} \label{sec:model}

We solved the vertically integrated hydrodynamics equations in cylindrical polar coordinates for a disc of ideal gas with surface density $\Sigma$, vertically integrated pressure $P$ and velocity $\vel=(u_r,u_\phi)$. We used the same equations  that are laid out in \cite{2019Kley}:
\begin{equation}
	\label{eq:navier-stokes}
	\begin{split}
	\DP{\Sigma}{t} & + \nabla\cdot(\Sigma\vel) = 0, \\
	\DP{\Sigma\vel}{t}& + \nabla\cdot\left(\Sigma\vel\otimes\vel + P \textbf{I} - \tensorGR{\Pi}\right)= \Sigma \bm{g}, \\
	\DP{e}{t}& + \nabla\cdot\left((e + P - \tensorGR{\Pi})\vel\right)=\Sigma\vel\cdot\bm{g} + Q. \\
	\end{split}
\end{equation}

Here, $\tensorGR{\Pi}$ denotes the viscous stress tensor, $\bm{g}$ refers to gravitational acceleration by the binary stars, and $Q$ encompasses any radiative terms.
For an ideal gas with adiabatic index $\gamma$, the isothermal and adiabatic sound speeds are related as $\ciso=\cs/\sqrt{\gamma}=\sqrt{P/\Sigma}=H\OmegaK$,
where $H$ is the gas pressure scale height and $\OmegaK=\sqrt{\G \Mco/r^3}$ the Keplerian angular velocity at distance $r$ from the central object(s) with mass $\Mco$. The gravitational constant is denoted with $\G$. The ideal gas relation also dictates that the specific internal energy of the gas is given by $e=P/(\gamma-1)$.
The temperature $T$ is then
\begin{equation}
T = \frac{\mu}{\Rgas} \ciso^2.
\end{equation}

Here, $\mu=2.353$ is the mean molecular weight and $\Rgas$ is the gas constant. These equations hold true for all models presented here.

\subsection{Temperature descriptions}

In this work we use three different prescriptions of radiative effects inside the disc: locally isothermal (no radiative effects), a parametrised cooling timescale, and self-consistent radiative cooling via thermal emission. In this section we describe each model in more detail.

\subsubsection{Locally isothermal model}
The locally isothermal prescription assumes that the temperature is a fixed radial profile, equivalent to a disc that responds instantly to heating or cooling and is always in thermal equilibrium. For a disc with an aspect ratio $h=H/r$, the temperature profile is given by
\begin{equation}
	\label{eq:temperature-aspect}
	T = \frac{\mu\G \Mco}{\Rgas} \frac{h^2}{r}.
\end{equation}
We choose a constant $h=0.05$, a value which is in good agreement with radiative simulations of binary systems by \cite{2019Kley}.

\subsubsection{Parametrised $\beta$-cooling}\label{sub:beta}

We use a $\beta$-cooling prescription similar to \citet{2001Gammie}, and follow the implementation detailed in \citet{2021Rometsch}. The $\beta$ relaxation model assumes that any temperature change is relaxed to a reference profile $T_0$ over a cooling timescale of $t_\mathrm{cool} = \beta\OmegaK^{-1}$ such that
\begin{equation}
	\frac{\partial T}{\partial t} = - \frac{T - T_0}{\beta}\OmegaK = \frac{Q_\text{relax}}{\Sigma\cv},\qquad \cv = \frac{\Rgas}{\mu(\gamma-1)},
\end{equation}
where $\cv$ is the heat capacity at constant volume. Assuming viscous heating with $Q_\text{visc}\approx \frac{9}{4}\nu\Sigma\OmegaK^2$ \citep{1978Tassoul} and the $\alpha$-viscosity model of \citet{1973alpha}, we have $\nu=\alpha\cs H$. Then, in a steady state with no compressional heating ($\nabla\cdot\vel=0$) it follows that
\begin{equation}
	\label{eq:equilibrium-temperature}
	Q_\text{visc} + Q_\text{relax} = 0 \Rightarrow T_\text{eq} = \frac{T_0}{1 - k\alpha\beta},\quad k:=\frac{9}{4}\sqrt{\gamma}(\gamma-1)
\end{equation}
with $k\approx 1.06$ for $\gamma=7/5$. In other words, the equilibrium temperature is higher for longer relaxation times or higher viscosity.

\subsubsection{Radiative models} \label{sub:rad-models}

For radiative models we use the thermal cooling prescription of \cite{2019Kley}. In a radiatively cooled system, the disc cools by emitting received heat as thermal radiation, the efficiency of which depends on the local environment. In these models, cooling is given by
\begin{equation}
	\frac{\partial \Sigma \cv T}{\partial t} = -2 \sigmaSB \frac{T^4}{\tau_\mathrm{eff}} = Q_\text{cool},
\end{equation}
where $\sigmaSB$ is the Stefan-Boltzmann constant. The effective optical depth $\tau_\text{eff}$ can be calculated with the emission optical depth $\tau$ following \cite{1990Hubeny} as
\begin{equation}
\tau_\mathrm{eff} = \frac{3}{8} \tau + \frac{\sqrt{3}}{4} + \frac{1}{4\tau+\tau_\mathrm{min}},\quad \tau=\int\limits_{0}^\infty\kappa_\text{R}\rho\dd z\approx\frac{c_1}{\sqrt{2\pi}}\kappa_\text{R}\Sigma.
\end{equation}
The coefficient $c_1=0.5$ is obtained by comparing 2D disc models with fully 3D calculations following \citet{2009Kley}.
In this equation $\kappa_\text{R}$ is the Rosseland mean opacity as modelled by \cite{1985Lin}, with the density and temperature power law relation detailed in \cite{2012Muller}. We use a minimum optical depth $\tau_\mathrm{min}=0.01$ to account for the extremely optically thin regions of the disc.

We can estimate the cooling timescale via this cooling prescription following previous studies \citep[e.g.][]{2020Miranda_I,2020Alex_II} by writing
\begin{equation}\label{eq:beta_rad}
	\DP{e}{t} \sim \frac{e}{t_\text{cool}} = |Q_\text{cool}| \Rightarrow \beta = \frac{\Sigma \cv T}{|Q_\text{cool}|}\OmegaK.
\end{equation}
Assuming thermal balance between viscous heating and thermal cooling, we can write that $|Q_\text{cool}| = Q_\text{visc}$ and find that $\alpha\beta=1/k$ or $\beta\approx1/\alpha$. As a result, we can estimate that in the absence of shock heating, the cooling timescale in our radiative model should be $\beta\approx10^3$. As we show later, this is indeed the case for the majority of the disc down to the cavity edge. Within the cavity region shock heating dominates due to proximity to the binary, while the very low gas density diminishes the efficiency of viscous heating which would modify the equilibrium state to $|Q_\text{cool}| = Q_\text{visc}  -P\nabla\cdot\vel$. For this reason, models where $\beta$ is constant run the risk of not treating cooling within the cavity region properly, especially for larger values of $\beta$. We address this issue in Sect.~\ref{sub:shift-beta}.

\begin{figure*}[t]
	\includegraphics[width=\linewidth]{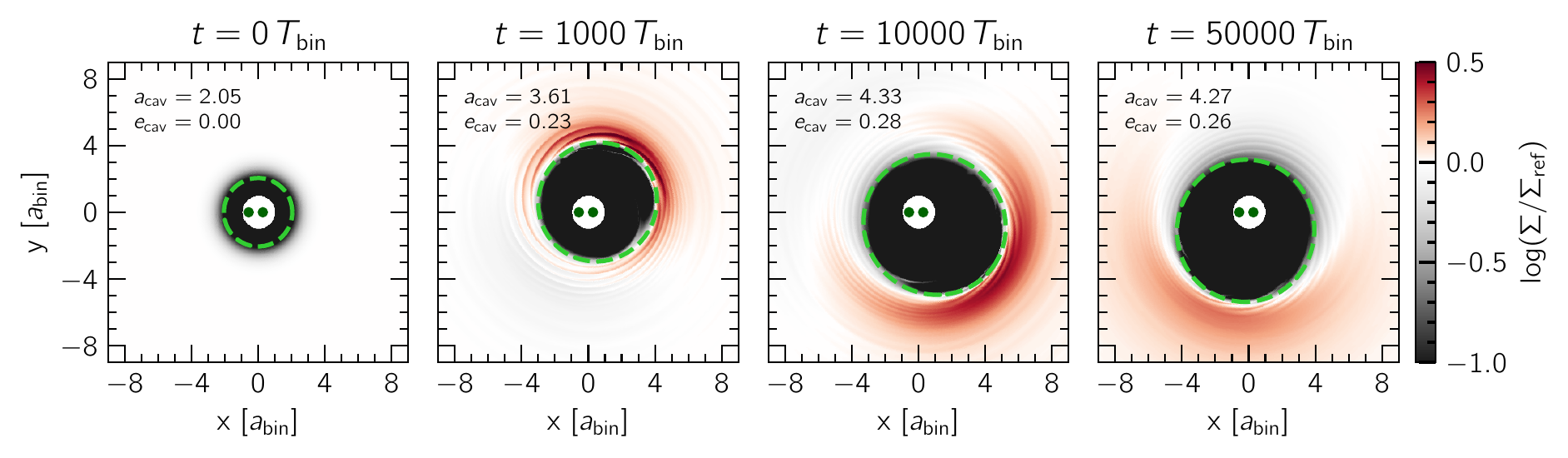}
	\caption{Time evolution of the 2D surface density normalised to its reference power-law profiles $\Sigma_\text{ref} \propto r^{-0.5}$ for the locally isothermal test case with $e\bin=0.15$. The system eventually reaches an equilibrium by $t = 50\,000~T_{\text{bin}}$. The dashed green curves represent the ellipse fitting. The binary spiral streams, which are present inside the cavity, are not visible within this colour range, but can be seen in Fig.~\ref{fig:shifting-beta-2d}.}
	\label{fig:iso-2d}
\end{figure*}

\subsection{Disc setup}\label{init}

We chose a binary setup that is comparable to the Kepler circumbinary planet-hosting systems discovered by the Kepler telescope. The total mass of the binary is $M_\mathrm{bin}=\Mco=1\,\Msun$, the binary semi-major axis is $a_\mathrm{bin}= 0.2\,\mathrm{au}$, and the mass ratio is $q=M_2/M_1=0.5$. We vary the binary eccentricity $e_\mathrm{bin} \in \{0.0, 0.15, 0.3, 0.5\}$ to cover both $e\bin$-dependent branches and the bifurcation point in the cavity shape for circumbinary discs found by \citet{2018Thun}. The binary does not receive feedback from the disc, maintaining a fixed orbit around its common centre of mass. The total initial mass of the gas is 1.7\% of $M_\mathrm{bin}$.

We initialise the disc with a circular cavity in the central region, such that the initial gas density is set to be
\begin{equation}
\Sigma_\mathrm{ini}=\Sigma_0 \left(\frac{r}{a_\mathrm{bin}}\right)^{-0.5} \cdot f_\mathrm{cav}, \quad f_\mathrm{cav} = \left[1 + \exp\left(-\frac{R-2.5\,a_\mathrm{bin}}{0.1\,a_\mathrm{bin}}\right)\right]^{-1},
\end{equation}
with $\Sigma_0=3743\,\text{g}/\text{cm}^2$.
The reference temperature profile corresponds to a constant $h=0.05$ as shown in Eq.~\eqref{eq:temperature-aspect}. For $\beta$ models, the initial temperature follows Eq.~\eqref{eq:equilibrium-temperature} directly such that the pressure is given by 
\begin{equation}
	P_\text{ini} = \frac{\Rgas T_\text{eq} \Sigma_\text{ini}}{\mu} .
\end{equation}
In all models we use $\alpha=10^{-3}$. 

We model the disc with the GPU-capable version of Pluto \citep{2017Thun, 2007Pluto} in a 2D cylindrical grid with $N_r \times N_\phi = 684 \times 1168$ cells, which corresponds to a resolution of $\approx9$ cells per scale height in both directions. The radial domain extends between 1--40\,$a_\mathrm{bin}$ with logarithmic spacing, thereby excluding the binary from the simulation domain for low eccentricities. However, this does not affect the structure of the disc in a significant way. We investigated this by lowering the inner radius in locally isothermal test runs and found that $1\,a_\mathrm{bin}$ offers accurate results while maintaining a reasonable timestep, in agreement with \cite{2021Tiede} who showed that material within inside $1\,a_\mathrm{bin}$ is accreted and does not change the dynamics of the outer disc.

At the inner boundary we implement a strict outflow boundary condition for density and radial velocity
 ($\partial_r\Sigma=0$, $v_{r\mathrm{,in}}=-|v_r(r_\mathrm{min})|$). 
The azimuthal velocity there is set to $\partial_r\Omega=0$ and the pressure is reflected at the inner boundary. At the outermost 10\% of the radial domain we damp $\Sigma$ and $P$ to their initial profiles and $u_r$ to zero. We use a density floor of $10^{-7}\,\Sigma_\mathrm{ini}$. In the radiative models we include a pressure limit equivalent to a maximum temperature of $3000\,\mathrm{K}$ and a minimum of $3\,\mathrm{K}$.

\section{Results} \label{sec:results}

In this section, we lay out the influence of cooling on the structure of the inner circumbinary disc with results from our numerical simulations. We first showcase the evolution of the circumbinary disc and the properties of its cavity with a locally isothermal model in Sect.~\ref{sec:test}. We then look at models with different cooling prescriptions in Sect.~\ref{sub:cools} and later analyse the effect of the parametrised cooling timescale $\beta$ on the disc shape by comparing to a model where $\beta$ drops over time in Sect.~\ref{sub:v-beta}. Finally, we demonstrate a simple model that can reproduce a temperature profile that is comparable to the radiative runs using $\beta$ as a function of $\Sigma$ in Sect.~\ref{sub:shift-beta}. 

\subsection{Example of disc evolution in the locally isothermal limit} \label{sec:test}

Most of the observed circumbinary planets orbit a binary with a moderate binary eccentricity $e\bin\leq0.21$.
Therefore, we choose simulations with a locally isothermal prescription, previously studied for Kepler binaries \citep[see][]{2017Thun, 2018Thun} with $e_{\text{bin}}=0.15$ as a standard of comparison for our models in the next section.

\begin{figure}[t]
	\includegraphics[width=\linewidth]{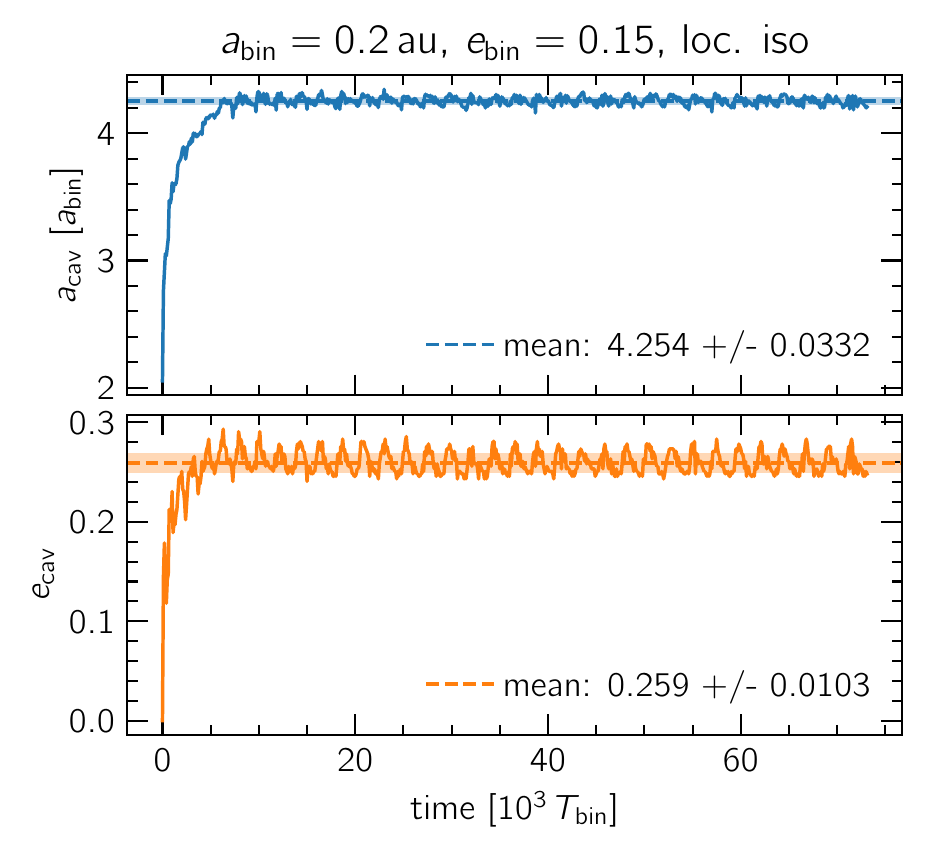}
	\caption{Time evolution of the semi-major axis and cavity eccentricity for the locally isothermal test case with $e\bin=0.15$. The dashed line represents the mean value around which the orbital parameters oscillate.}
	\label{fig:iso-timeevo}
\end{figure}

 In Fig.~\ref{fig:iso-2d} we display the disc evolution in 2D snapshots at $t=\{0,1000,5000,50\,000\}\,T\bin$. The binary clears out an eccentric inner cavity in the disc that eventually stabilises to an ellipse with a semi-major axis of $a_\text{cav}=4.25\,a\bin$ and an eccentricity $e_\text{cav}=0.26$, as can also be seen in Fig.~\ref{fig:iso-timeevo}.

In all 2D plots, a dashed green bounded curve traces an ellipse that is fit to the edge of the cavity, which is defined as the region where $\Sigma$ drops by a factor of 10\% compared to the current azimuthal density maximum. This is done similarly to \cite{2017Thun}, using the peak density location to compute the argument of the ellipse's apocenter and constructing the ellipse by finding the cavity edges along the line connecting its apsides, using the centre of mass as the focal point.

In Fig.~\ref{fig:iso-timeevo} the orbital parameters of the cavity oscillate around a mean value. This oscillation is caused by the precession of the disc around the binary. Especially for large $e\bin$, the cavity shape experiences large oscillations caused by the precessing motion of the CBD around the binary and the changing alignment between the argument of periastron of the disc and the binary orbits. At times of minimal cavity size and eccentricity, the disc and the binary orbits are in alignment. \cite{2018Thun} have shown that the precession is directly linked to the cavity shape in the locally isothermal case by relating it to the theoretical precession of a massless third body. We discuss this in the context of our models in the next section.

\begin{figure*}[t]
	\includegraphics[width=\linewidth]{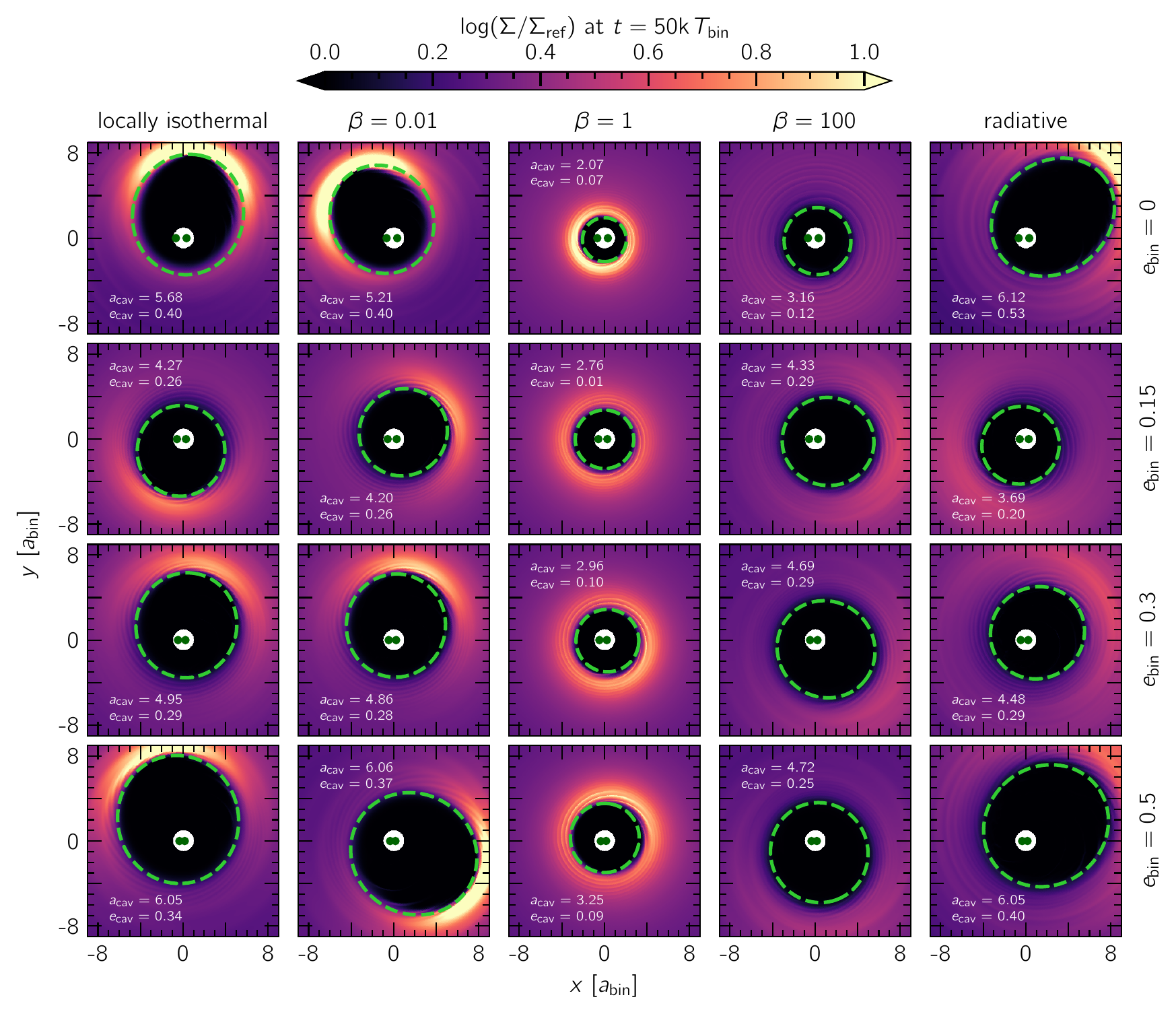}
	\caption{Comparison of the 2D surface density at steady state of all models in Table~\ref{tab:models} after $50\,000\,T\bin$. The surface density is normalised to the reference power-law profile ($\Sigma_\text{ref}\propto r^{-0.5}$). The annotated values for $a\cav$ (in $a\bin$) and $e\cav$ refer to the ellipse fit in dashed green lines.}
	\label{fig:equilibria-2d}
\end{figure*}

\subsection{Comparison of cooling models} \label{sub:cools} 

In the previous section we analysed a sample circumbinary disc model as it evolves towards a quasi-steady state and laid out our method of measuring the cavity parameters in equilibrium with our locally isothermal example model. Here, we investigate a range of models with different $e\bin$ and cooling prescriptions as listed in Table~\ref{tab:models} using this method. Each model has run for at least $50\,000\,T\bin$, until the cavity profile has converged.
\begin{table}[b]
	\centering
\begin{tabular}{|c|c|c|c|c|c|}
\hline 
$e\bin$ & iso & $\beta=10^{-2}$&$\beta=1$&$\beta=10^{2}$ & radiative \\ 
\hline 
0.0 & >5.68 &5.26 & 2.12& 3.06& 6.73\\ 
    & 0.43 &0.41 & 0.08 &0.11& 0.57\\
\hline 
0.15 & 4.25& 4.20& 2.76& 4.35& 3.70 \\ 
     & 0.26& 0.26& 0.05& 0.28& 0.23\\ 
\hline 
0.3 &  4.94 & 4.89& 3.02& 4.61&4.36\\ 
    &  0.29 & 0.30& 0.07& 0.27&0.28\\ 
\hline 
0.5 &  6.07 & 5.99& 3.15& 4.78&6.05\\ 
&  0.35 & 0.35& 0.08& 0.25&0.39\\ 
\hline 
\end{tabular}
\caption{List of cavity semi-major axis $a_\text{cav}$ (top, quoted in $a_\text{bin}$) and eccentricity $e_\text{cav}$ (bottom) of models discussed in Sect.~\ref{sub:cools}. A '$>$' symbol implies that the profile has not fully converged (see Fig.~\ref{fig:time-evo-all} for details). }
\label{tab:models}
\end{table}

Fig.~\ref{fig:equilibria-2d} shows 2D snapshots of the disc surface density at a time of $50\,000\,T\bin$. The cavity becomes small and circular for $\beta=1$ independent of $e\bin$, and its size grows with binary eccentricity. The latter is in agreement with the instability region depending on binary eccentricity as seen in the early dynamical analysis by \cite{1994Arty} or the numerically evaluated instability region by \cite{1999Holman}, giving a minimum stable region for the disc or planets. This is a hint that the conditions for the $\beta=1$ case lead to reduced hydrodynamical interaction inside the disc.
Radiative and locally isothermal models  produce larger and more eccentric cavities in all cases, as they correspond to an effective $\beta\approx10^3$ and $\beta\rightarrow0$ respectively. The high- and low-$\beta$ models are accordingly comparable to the radiative or locally isothermal case with the exception of $\{e\bin=0,\,\beta=100\}$, where the disc remains circular, while the locally isothermal and radiative models show very eccentric cavities. To a lesser degree this also affects the highly eccentric disc caused by the eccentric binary in our model with $\{e\bin=0.5,\,\beta=100\}$. The possible causes and solutions for problems with the $\beta=100$ cases will be addressed in Sect.~\ref{sub:shift-beta}. The cavity size and eccentricity in radiative and isothermal models are smallest for $e\bin=0.15$, in agreement with \cite{2018Thun}. For very eccentric binaries ($e\bin=0.5$) the cavity becomes very wide and eccentric, which also leads to wider planetary parking orbits around such systems as is the case for Kepler-34b \citep{2012Kepler34-35, 2021Penzlin}.

\begin{figure*}[h]
	\includegraphics[width=\linewidth]{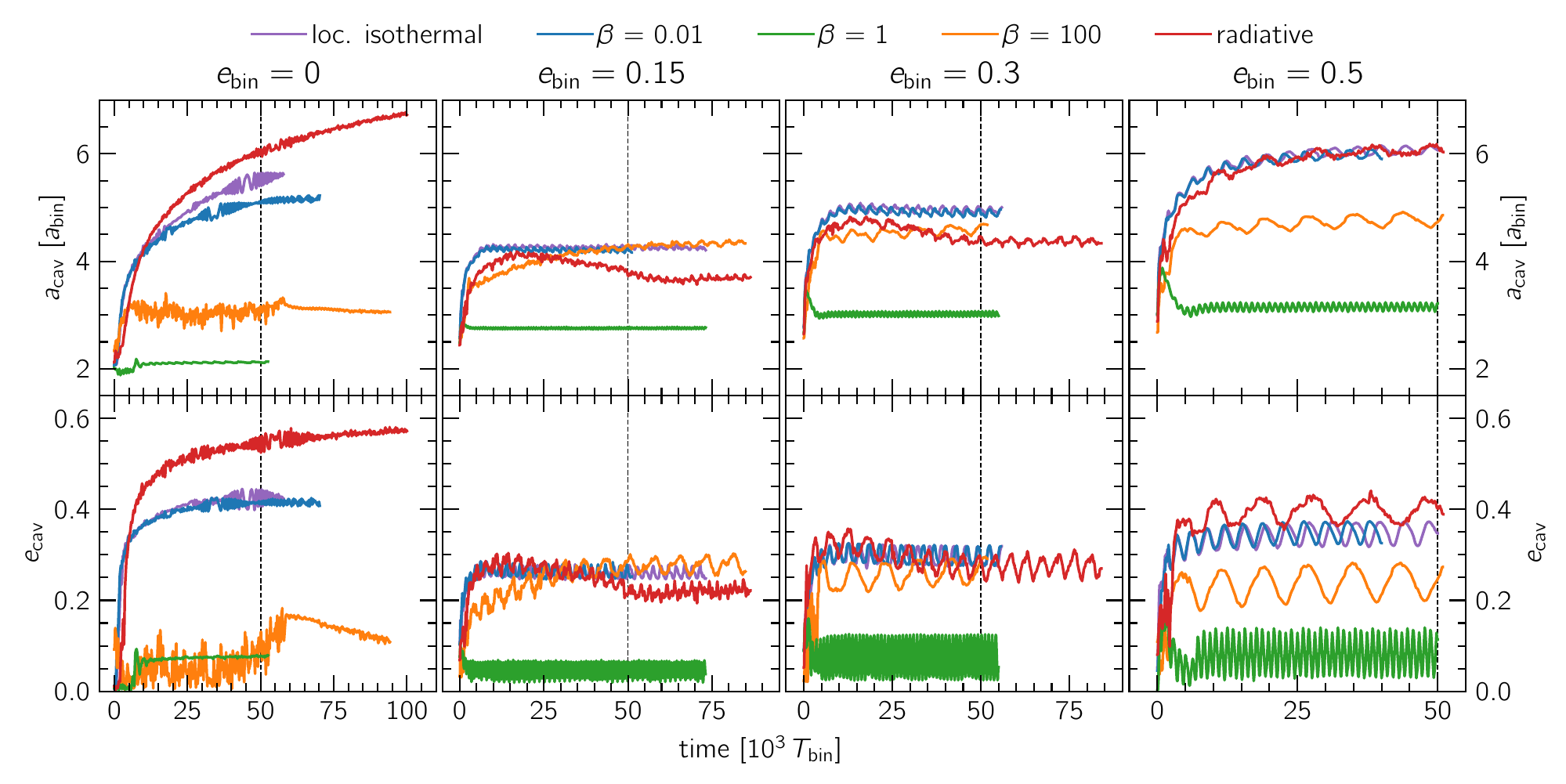}
	\caption{Time evolution of $a_\text{cav}$ and $e_\text{cav}$ for all models presented in Table~\ref{tab:models}. The dashed vertical line marks the timestamp at $t = 50\,000~T_{\text{bin}}$, which is used for comparison in Fig.~\ref{fig:equilibria-2d}. The data shown has been smoothed with a rolling mean over $300~T_{\text{bin}} $ for a clearer view.}
	\label{fig:time-evo-all}
\end{figure*}

The time evolution of the cavity parameters $a\cav$ and $e\cav$ is shown in Fig.~\ref{fig:time-evo-all}. The average values of the cavity shape over the last $10\,000\,T\bin$ are listed in Table~\ref{tab:models} and can differ slightly from the values at the time of the snapshot in Fig.~\ref{fig:equilibria-2d}. The curves in Fig.~\ref{fig:time-evo-all} show oscillations around these average values, an effect caused by the precession of the disc. As visible in the oscillations in Fig.~\ref{fig:time-evo-all}, typical precession periods are in the order of $\sim 10^3\,T\bin$ \citep[see also][]{2017Thun}. The oscillation amplitude increases with higher $e\bin$, as the disc pericenter approaches the binary apocenter more closely once per precession. 

The cavity in the radiative model with $e\bin=0.15$ reaches its maximum size after $20\,000\,T\bin$ but continues evolving, shrinking to slightly lower final values of $a\cav$ and $e\cav$ over $50\,000\,T\bin$ as the disc surface density and temperature profiles readjust to an equilibrium state that corresponds to a smaller but self-consistently calculated aspect ratio $h\approx0.04\text{--}0.05$ in the inner disc \citep[also shown in][]{2019Kley}. Given this slightly lower $h$, a smaller and less eccentric cavity is to be expected (Penzlin et al., in prep.). 

Contrary to other models, where the cavity reaches an equilibrium state within $\approx50\text{k}\,T\bin$, the cavity of the radiative model with $e\bin=0$ is still increasing in size even after $>100\,000\,T\bin$. While it is not yet fully understood why this is the case, we speculate that it might be the result of a combination of a very long effective cooling timescale $\beta\approx 10^3$ (see Sect.~\ref{sub:rad-models}) with the generally longer evolution timescale of models with $e\bin=0$ due to the accumulation of long-lived eccentricity modes created by the symmetrically rotating binary that are trapped by the steeply truncated disc \citep{2020Munoz_theo}.

\begin{figure}[h]
	\includegraphics[width=\linewidth]{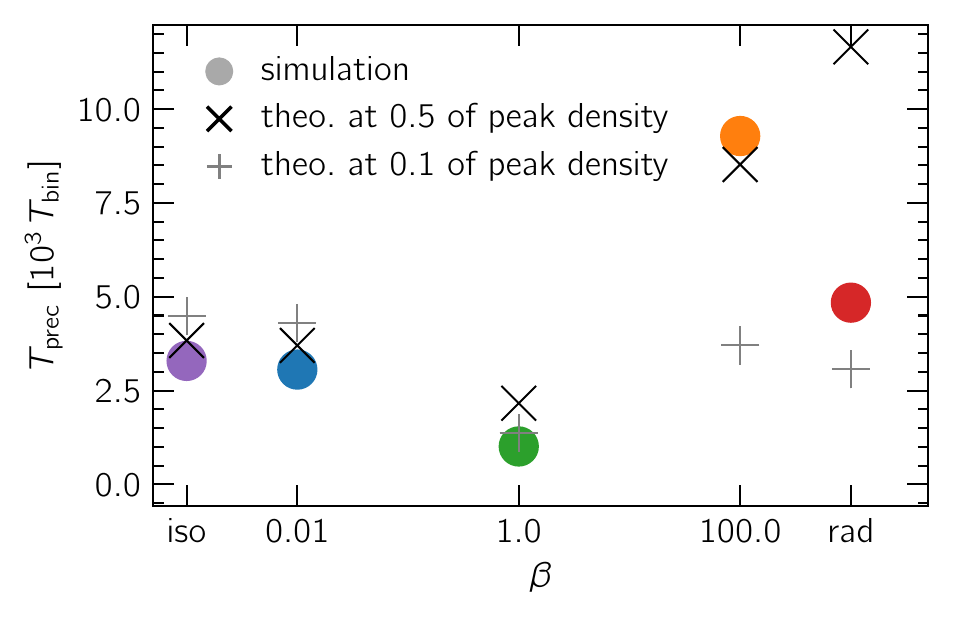}
	\caption{Precession period of the disc (coloured dots) for all models with $e\bin=0.3$. For comparison, the 3-body precession period for the 10\% and 50\% peak density ellipse are plotted with "+" and "$\times$". For the 3-body precession, the semi-major axis and eccentricity have been averaged over $4000\,T\bin$.}
	\label{fig:tprec}
\end{figure}

The precession rate of the discs is of the order of a few $1000\,T\bin$ and the exact number for the models with $e\bin=0.3$ derived from the argument of periastron of the inner disc is displayed in Fig.~\ref{fig:tprec} (coloured dots). When the density peaks are narrow as Fig.~\ref{fig:equilibria-2d} shows, this precession can be approximated with a three-body orbit at the cavity edge for locally isothermal models like in \cite{2018Thun} 
\begin{equation}\label{eq:t-prec}
	T_\mathrm{prec} = \frac{4}{3} \frac{(q+1)^2}{q} \left(\frac{a\cav}{a\bin}\right)^{3.5} \frac{(1-e\cav)^2}{1+1.5 e_\mathrm{bin}^2} T_\mathrm{bin},\quad q=M_2/M_1.
\end{equation}
However, in discs with a wider density maximum like in the high $\beta$ or radiative cases \citep{2019Kley}, the period of the precession increases as the disc precesses rigidly and the associated third-body orbit shifts outwards. Therefore, even for equal cavity sizes between radiative and locally isothermal models, the precession rate linked to the width and position of the peak density can still differ, and is usually higher for longer cooling timescales. This can be understood by comparing the precession rates of the simulations with those obtained from Eq.~\eqref{eq:t-prec} for ellipses that are defined at locations in the cavity where the density is 10\% and 50\% of the peak density, as shown in Fig.~\ref{fig:tprec}. For locally isothermal models and those with $\beta\leq1$ the density peak is so narrow that both values match the precession well. However, for $\beta=100$ the density peak is much wider so that the 3-body precession for the 10\% case is significantly lower and does not match the precession rate anymore. This difference in peak position and precession rate is a result of the higher pressure in the $\beta=100$ case near the inner edge, which influences the propagation of excitations in the disc similar to the effect studied in \cite{2016Teyssandier}. The radiative model behaves differently, showing a minor increase in precession period. This is likely caused by the reduced pressure in the inner disc \citep{2019Kley}, leading to less rigid disc rotation that affects a smaller section of the disc.

\subsection{The influence of $\beta$ on the cavity shape} \label{sub:v-beta}

\begin{figure}[h]
	\includegraphics[width=\linewidth]{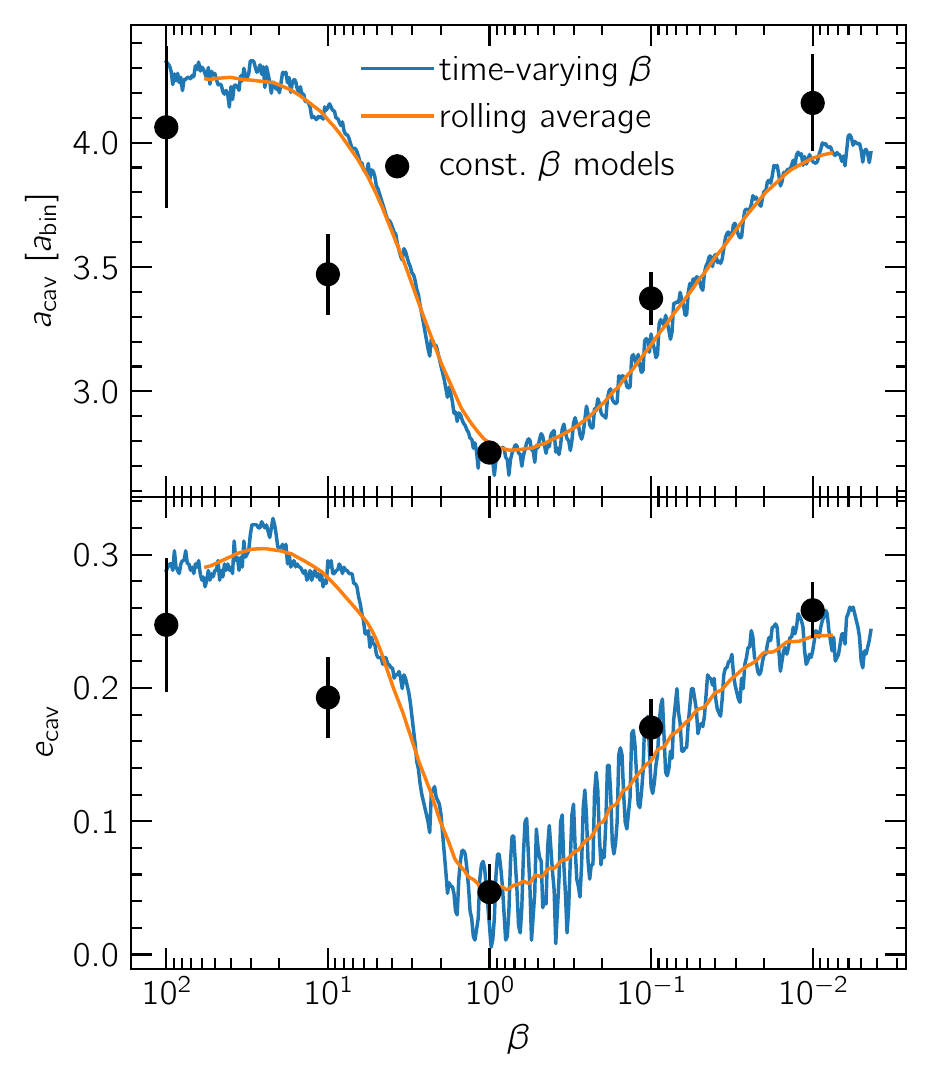}
	\caption{Cavity eccentricity and semi-major axis of the models with $e\bin=0.15$ and constant $\beta$ values (black dots) and the model with time-varying $\beta$ (blue curve) following Eq.~\eqref{eq:beta-over-time}. The orange curve is the rolling average of the blue model over $5000\,T\bin$.}
	\label{fig:v-shape}
\end{figure}

As we have seen in Sect.~\ref{sub:cools}, the cavity for $\beta=1$ becomes small and nearly circular, while generally circumbinary disc cavities are eccentric even for large variations in aspect ratio and viscosity for locally isothermal \citep[see e.g.][]{2017Thun, 2020Ragusa, 2020Munoz} as well as radiative discs \citep{2019Kley}. We ran models for $e\bin=0.15$ with $\beta=\{10^{-1},10^1\}$ in addition to the values in Table~\ref{tab:models} and calculated the converged cavity shape for all five $\beta$ models. The resulting V-shape, displayed by the points in Fig.~\ref{fig:v-shape}, is centred around $\beta=1$ and verifies that this value most likely corresponds to a total minimum.

Based on our results using different models with a constant $\beta$, we constructed a model of a disc where $\beta$ varies in time from $10^2$ to $10^{-2}$ over $40\,000\,T\bin$ with
\begin{equation}
	\label{eq:beta-over-time}
	\log\beta = 2 - \frac{t}{10^4\,T\bin},
\end{equation}
using the equilibrium state of the $e\bin=0.15$, $\beta=10^2$ model at $50\,000\, T\bin$ as our initial state. The evolution of the cavity shape for this model is displayed in Fig.~\ref{fig:v-shape}. The averaged curves show a clear minimum at $\beta=1$ for both $a\cav$ and $e\cav$, with the cavity shrinking and circularising as $\beta$ approaches this value. Due to the initially long cooling timescale, the disc's response to the cavity-shrinking effect as $\beta$ approaches unity is significantly slowed, so that the orange curve does not exactly match our data points during its descent towards $\beta=1$. As a result, $e\cav$ is higher than the value predicted for our fixed-$\beta$ model when $\beta(t)\approx10$. For $\beta\lessapprox1$, however, our variable-$\beta$ model agrees very well with our models with a constant $\beta$.

The periodic variations in the blue curve in that figure are an artefact of changes during one precession period of the disc around the binary, as analysed in \cite{2017Thun}, \citet{2018Thun} and \citet{2019Kley} (see also Fig.~\ref{fig:tprec}). Therefore, even though the cavity sizes at both ends of the curve are comparable, the precession period varies due to the difference in the inner disc structure as described in Sect.~\ref{sub:cools}.

\begin{figure*}[t]
	\includegraphics[width=\linewidth]{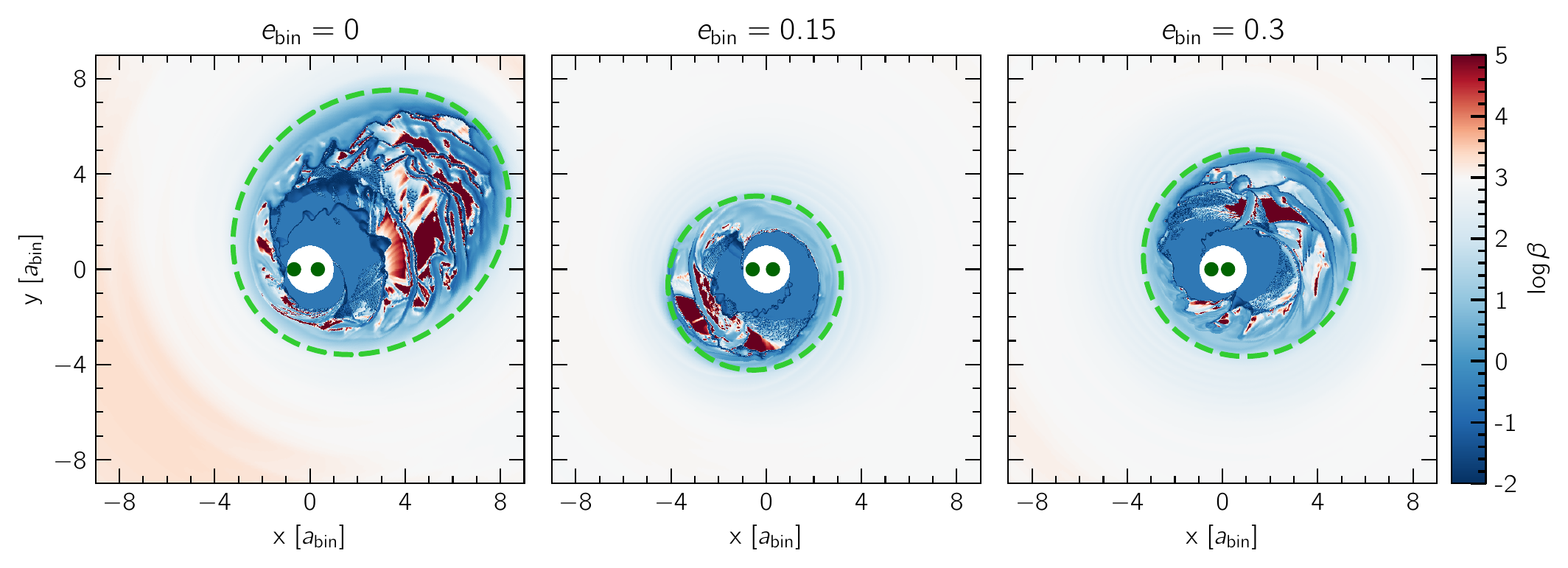}
	\caption{Heatmaps of $\beta$ in radiative models. The cavity is marked by the green dashed line. We highlight the uniformly shaded disk, indicating thermal equilibrium between viscous heating and $\beta$-cooling, and the shock-dominated cavity where $\beta$ deviates significantly from the expected value of $10^3$. Regions with a very low $\beta$ inside the cavity act as an indicator of spiral streamers, where heating is strongly dominated by shocks.}
	\label{fig:shifting-beta-2d}
\end{figure*}

\begin{figure}
	\includegraphics[width=\linewidth]{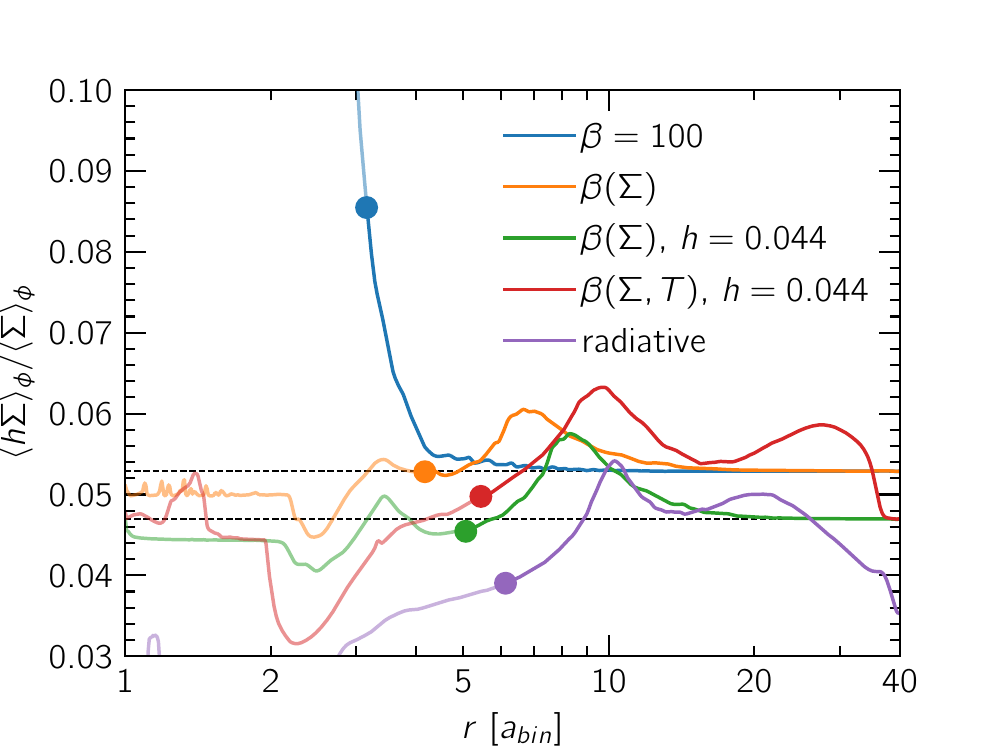}
	\caption{Comparison of azimuthally averaged, density-weighted aspect ratio profiles of the fiducial model with $\beta = 100$ and $h=0.05$ (blue) against density-dependent $\beta(\Sigma)$ for $h=0.05$ (orange) and $h=0.044$ (green), adaptive $\beta(\Sigma,T)$ according to Eq.~\eqref{eq:beta_rad} (red), and radiative models (purple) after $50\,000\,T_{\text{bin}}$ for $e_{\text{bin}} = 0$. Horizontal dashed lines denote the equilibrium state for $\beta$ models where shock heating would be absent. Coloured dots mark the respective $a_{\text{cav}}$.}
	\label{fig:example-aspect}
\end{figure}

\begin{figure}
	\includegraphics[width=\linewidth]{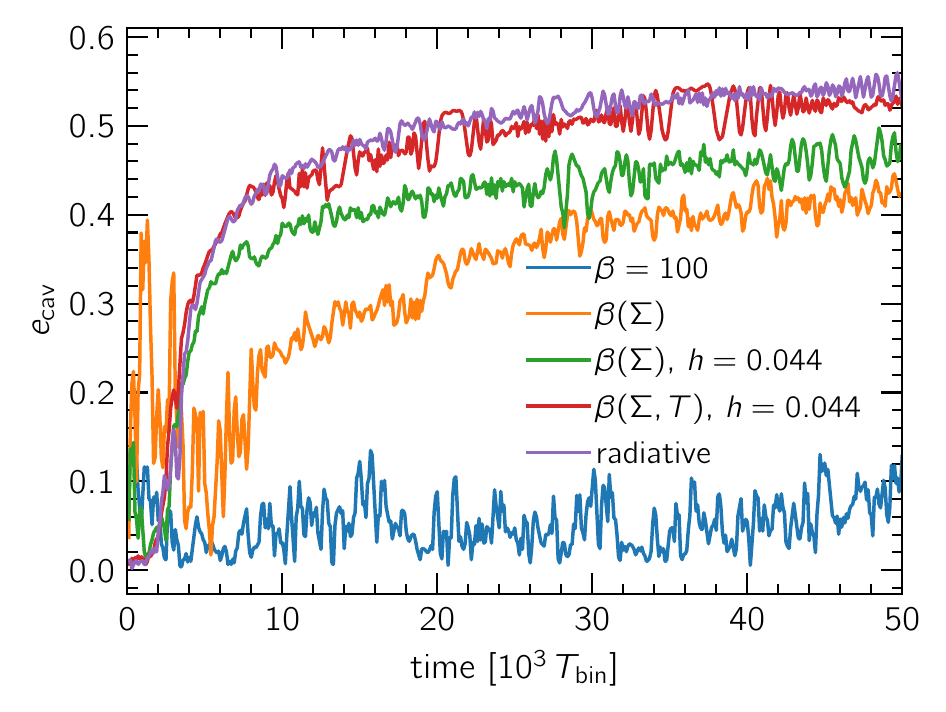}
	\caption{Time evolution of $e_{\text{cav}}$ for the models shown in Fig.~\ref{fig:example-aspect}. The data shown has been smoothed with a rolling mean over $300~T_{\text{bin}} $ for clarity.}
	\label{fig:adj-evol}
\end{figure}

\subsection{Accounting for the cavity in models with slow cooling} \label{sub:shift-beta}

In the models presented, the actual pressure scale height can vary from the initially prescribed profile as the disc equilibrium temperature is changed by spiral shock heating or radiative cooling. This also affects models with $\beta \gg 1$ significantly, as the slow cooling cannot counteract spiral shock heating efficiently near the binary. The effective $\beta$ of radiatively cooled models, computed using Eq.~\eqref{eq:beta_rad}, is displayed in Fig.~\ref{fig:shifting-beta-2d}. As described in Sect.~\ref{sub:rad-models}, it corresponds to $\beta\approx 10^3$ inside the disc, but drops significantly within the cavity with the exception of spikes due to a combination of viscous and spiral shock heating around very low density streamers. This indicates that viscous heating is not efficient enough within the cavity, consistent with previous works \citep{2019Kley}, and suggests that the cooling timescale should be much lower in this low-density region leading to unnaturally increased temperatures for the high-$\beta$ models near the cavity.

\begin{figure*}[t]
	\includegraphics[width=\linewidth]{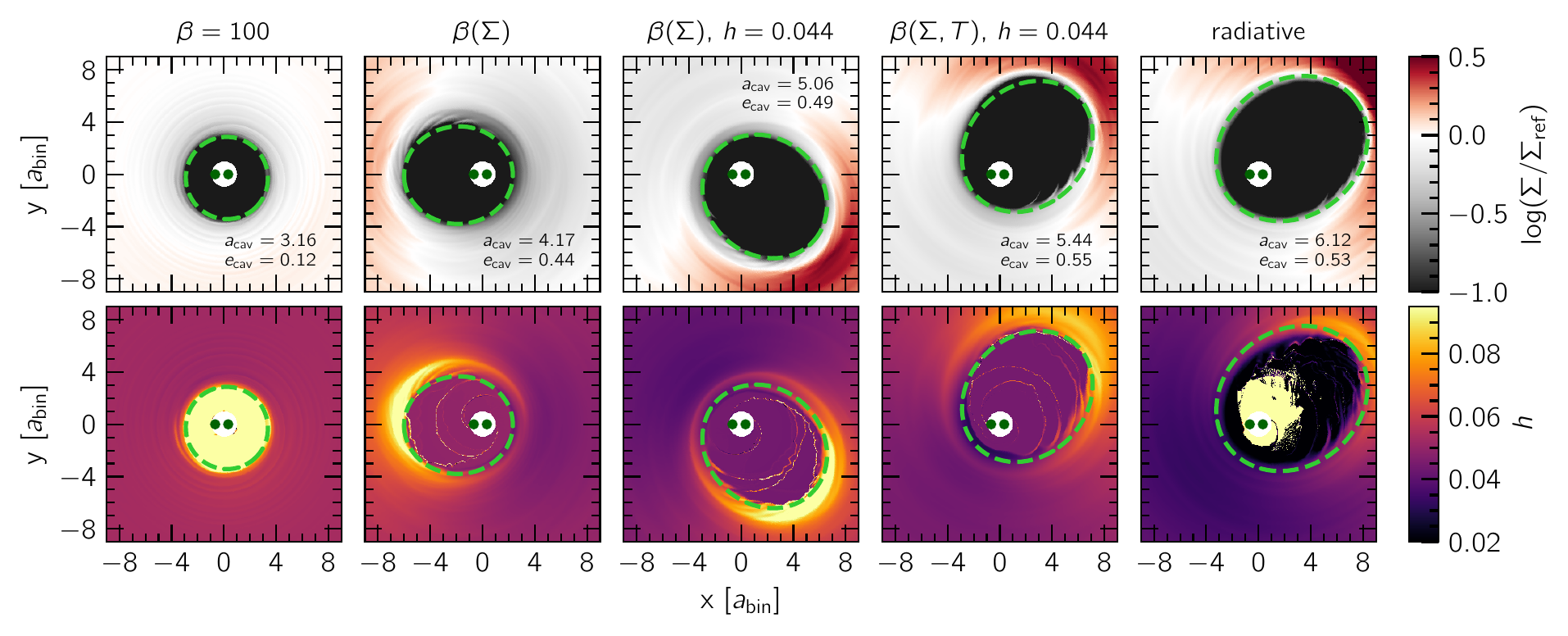}
	\caption{Comparisons of surface density (top) and aspect ratio (bottom) heatmaps between (from left to right) $\beta = 100$, density-dependent $\beta(\Sigma)$ ($h=0.05$ and $h=0.044$), adaptive $\beta(\Sigma,T)$ and radiative models for $e\bin=0$, at time $50\,000~T\bin$. The top panels have been normalised to their reference power-law profiles $\Sigma_\text{ref} \propto r^{-0.5}$.}
	\label{fig:adjusted-2d}
\end{figure*}

As already seen in Sect.~\ref{sub:cools}, some models with $\beta=100$ display unexpectedly circular cavity structures. 
To understand the actual disc properties of the models which influence the dynamics, we plotted the aspect ratio $h$ for the $\beta=100$ model and the radiative model with $e\bin=0$ in Fig.~\ref{fig:example-aspect}. The aspect ratio here is derived from the disc with 
$h = \cs r/\Omega_\mathrm{K}$
as shown in Sect.~\ref{sec:model}. Additional models that are described further below aim to reconcile the differences between the two main runs. Coloured dots mark the radial distance where the density is lower than $10\%$ of the peak density, which is the criterion we use to define the cavity edge.
High-$\beta$ regions inside the cavity as seen in Fig.~\ref{fig:shifting-beta-2d} contain so little material that they show little relevance on the density-weighted $h$.
The radiative model has a nearly constant, low $h$ within the disc reaching into the cavity. In radiative models, the aspect ratio spikes only close to the binary due to shocks created by the binary motion as also seen in Fig.~\ref{fig:shifting-beta-2d}. For that reason, density weighting was used to remove spikes due to shock heating inside the low density region in Fig.~\ref{fig:example-aspect}. The azimuthal dependence of $h$ is better highlighted in Fig.~\ref{fig:adjusted-2d}.

However, for the $\beta=100$ model the aspect ratio increases substantially within the disk near the cavity edge. The very high aspect ratio ($h>0.1$) of the $\beta=100$ model at the inner edge leads to a stronger local viscosity ($\nu\propto h^2\sqrt{r}$), which can suppress the excitation of the eccentric cavity by smearing out the wakes in the cavity caused by the binary motion while also refilling the forming cavity more efficiently. This hints at the temperature profile inside the cavity playing an important role in the final shape of said cavity.

In light of this deviation of the aspect ratio profiles, we implemented two additional models. First, a density-dependent prescription $\beta(\Sigma)$ where $\beta$ is constant within the disc but drops in low density regions, when the density drops below the threshold value $\Sigma_\mathrm{th}=5\%\,\Sigma_0$. Then, for a given input $\beta_\text{ref}$, the density-dependent $\beta(\Sigma)$ is given as follows:
\begin{equation}
\log\beta(\Sigma) = \log\beta_\text{ref} - (2+\log\beta_\text{ref}) \cdot \frac{\Sigma_\mathrm{th}-\Sigma}{\Sigma_\mathrm{th}}, \quad \Sigma\leq\Sigma_\mathrm{th},
\end{equation}
such that it drops smoothly towards 0.01 within the cavity. The resulting aspect ratio profile of this model is given by the orange curve in Fig.~\ref{fig:example-aspect} and avoids the unrealistic long cooling time near that cavity edge that leads to the unphysical increase in temperature in the inner disc.

The second additional model uses instead an adaptive $\beta(\Sigma,T)$ that depends on the local thermodynamical properties of the gas following Eq.~\eqref{eq:beta_rad}, while relaxing the temperature profile to one where $h=0.044$ in order to mimic the radiative model more closely. To stabilise this method in the cavity, where $\beta$ varies strongly with position, as well as to prevent runaway heating when $\beta>1/\alpha$ (see Sec.~\ref{sub:rad-models}), we limit the value of $\beta$ between $10^{-3}$ (locally isothermal limit) and $10^3$ (adiabatic limit). This prescription, which attempts to emulate the results of the radiative model, is shown with red curves in Figs.~\ref{fig:example-aspect}~and~\ref{fig:adj-evol} and shows very good agreement with the radiative model in terms of cavity properties. 
This model also closely follows the development of $e\cav(t)$ in the radiative model as can be seen on Fig.~\ref{fig:adj-evol}, while the simpler density-dependent $\beta(\Sigma)$ models both lag behind the radiative evolution. Nevertheless, $e\cav$ in all adaptive $\beta$ models follows an increasing trend and is closer to the radiative model than the locally isothermal model discussed in Sec.~\ref{sub:cools}.

The heatmap in Fig.~\ref{fig:adjusted-2d} also shows that the structure and size of the inner cavity of the improved models are comparable, while the simple $\beta=100$ model deviates significantly. The plotted maps of $h(r,\phi)$ illustrate this behaviour, with the $\beta=100$ model having a substantially higher aspect ratio in the cavity. The uniformly high pressure in the cavity for $\beta=100$ suppresses the wakes launched by the binaries that are otherwise visible in all other models by streams of high aspect ratio caused by shocks inside the cavity.

However, the adjusted $\beta(\Sigma,T)$ model requires a priori knowledge of the thermal profile of the disc, while at the same time results in a strictly hotter disc in regions where shock heating is negligible due to the very long cooling timescale in the bulk of the disc (see Fig.~\ref{fig:shifting-beta-2d} and Eq.~\eqref{eq:equilibrium-temperature}). This excess heating becomes evident by looking at the aspect ratio profile of this model in Figs.~\ref{fig:example-aspect}~and~\ref{fig:adjusted-2d}, which show that this adaptive model behaves similarly to the radiative model outside of the cavity but its $h$ is consistently higher in value.

\section{Discussion}
\label{sec:discussion}

In this section we consider the influence of different physical effects as well as binary parameters on the topic of the applicability of our results on observed systems. We also highlight some caveats of our models and discuss possible improvements in future modelling, with particular focus on larger circumbinary discs.

\subsection{Effective range of thermal models}
%Difference to large CBD + irradiation heating
In this study we focus on configurations that represent the circumbinary planet systems observed by the Kepler mission, and therefore we chose close binary configurations with $a\bin = 0.2\,\mathrm{au}$. In this regime, viscous heating is the dominant heat source as also shown in \cite{2019Kley}, while stellar irradiation has a negligibly small contribution. Assuming thermal equilibrium, the cooling time inside the disc in this regime is then $\beta\approx1/\alpha$, as shown in Sect.~\ref{sub:shift-beta}. Therefore, the disc of these close-in systems can be reasonably well-represented by density-dependent $\beta(\Sigma)$ models with $\beta_\text{ref}\gtrsim10^2$ if one takes into account that the cooling timescale is much shorter within the cavity and that the aspect ratio in radiative models can deviate towards slightly lower values in thermal equilibrium.

For very large discs such as GG~Tau with a cavity size of $a\cav\sim180\,\mathrm{au}$, irradiation heating \citep[e.g.][]{1997ChiangGoldreich,2004MenouGoodman} would become more relevant compared to viscous heating, which would result in lower temperatures and conditions closer to locally isothermal as the disc becomes optically thinner. While $\beta$ models are easily scalable, viscously heated models only represent the circumbinary disc around close binaries. To investigate realistic thermal conditions around observed large CBDs, a model should account for stellar irradiation in addition to viscous heating. To adapt a $\beta$ model to such a more realistic case, $\beta$ would need to depend on radius or even azimuth, as well as drop in optically thin regions such as inside the cavity or at large radii, far from the binary. In that sense, a lower $\beta$ corresponds to the case of an even wider disc. As a result, the different cases introduced in this work bear relevance to wide circumbinary discs, especially ones where locally isothermal conditions are likely. Models including irradiative terms would be the topic of future work. 

However, as we have demonstrated, the main difference between our radiative and locally isothermal models even for short scales is the precession rate, while conditions close to a $\beta=1$ case cause the disc to radically change its cavity structure. This effect may be exacerbated in this work as our models do not include the effects of in-plane cooling via flux-limited diffusion \citep{2020Miranda_II}, which could possibly mediate the interaction between cavity and disc in this case. This could be investigated further in future work.

\subsection{Planet formation scenario}
The models presented here are created with the close-in circumbinary planets in mind. We cover the full range of binary eccentricities relevant to the planet case from circular binaries like Kepler-47 \citep{2019Kepler47} and Kepler-413 \citep{2014Kepler413} to the most eccentric case of Kepler-34 \citep{2012Kepler34-35}. Such discs are a representation of environments of planet formation around binaries. However, given the masses of circumbinary planets of $\sim 0.1 M_\mathrm{Jup}$, an inserted planet will change the dynamics of the disc \citep{2021Penzlin}. Especially when it reaches an orbit inside the cavity, the planet will shield the disc from further excitation by the binary, leading to a circularisation that may appear similar to the $\beta=1$ case, albeit caused by different interactions.

%Effects of M2/M1
\subsection{On the binary mass ratio}
Our simulations all use a mass ratio of $M_1/M_2 = 0.5$. Using locally isothermal models,
\cite{2018Thun} have shown that variations of the mass ratio only have a minor effect on the cavity shape. \cite{2016q-transition} found a behaviour transition in discs around binaries with mass ratios above $M_1/M_2 > 0.04$ that leads to asymmetric, time-variable discs with a large cavity. As a result, for stars of comparable size the large, eccentric cavities explored in this work are expected. We chose an intermediate mass ratio of 1:2 such that our general results can be expected to be comparable to binaries within a range of mass ratios. The exact shape and properties of the cavity as a function of the binary mass ratio is beyond the scope of this study.

\subsection{Three-dimensional effects}
We use a 2D cylindrical grid. As our simulations require up to $10^5\,T\bin$ for the cavity to converge to a steady state, running 3D models is simply not feasible. \cite{2012Shi} performed 3D MHD simulations of a circumbinary disc for up to $1000~T\bin$, and found an increasing trend in disc eccentricity. \cite{2020Pierens} showed the same trend in their 3D simulations of CBDs, where comparably eccentric cavities develop. Using 3D smooth particle hydrodynamics, \cite{2020Ragusa} also reproduced the growth of eccentric cavities for locally isothermal models. Hence, neglecting the vertical direction in our simulations does not appear to affect the general picture of an eccentric circumbinary disc with a large central cavity. One drawback of 3D simulations is that the computational cost does not allow for the long simulation times necessary to reach a converged state with the precessing disc, as can be seen in our 2D models. As the cavity precesses, its properties vary as shown in Fig.~\ref{fig:iso-timeevo}. As precession times can be in the order of $10^3~T\bin$ it is difficult to assess the convergence between 2D models that run for $50\,000~T\bin$ and 3D models that run for only $1000\text{--}3000~T\bin$. The eccentricity reached at the end of the aforementioned 3D simulations is compatible with that of our 2D simulations at the same time.
Nevertheless, the exact structure and properties of the cavity could be different in a fully 3D model in equilibrium. Such a model is beyond the scope of this study.

\subsection{Computational domain}

In this study we chose an inner domain boundary at $1~a\bin$ such that the full interaction of the binary is not included for runtime reasons. \cite{2021Tiede} shows that inside this radius nearly all mass is accreted and does not alter the dynamics of the outer disc. 
As Fig.~\ref{fig:shifting-beta-2d} also reveals, the mass of the streamers inside the cavity is 3 to 4 orders of magnitude smaller than the disc density at the converged state for the chosen viscosity. In very viscous or thick systems for example, advection can become the more relevant term for angular momentum transport and can play an important role in the dynamics of such discs \citep[e.g.][]{2012Shi, 2019Munoz}.
In these protoplanetary disc cases the gravitational torque drives the eccentricity of the disc as described in \cite{2020Munoz_theo}.
At the same time, the effect of the disc on the binary motion is not taken into account in our model. For the chosen viscosity the orbital change of the binary is in the order of $10^{-7}~a\bin/T\bin$ \citep{2022Penzlin} and thereby much longer than the relevant timescales of the simulation.

\subsection{Different cooling prescriptions}

As mentioned in Sect.~\ref{sub:shift-beta}, we implemented additional models with $\beta$-cooling in an effort to construct a prescription that can reproduce the results of our radiative runs while also gaining the speed advantage of $\beta$ methods, which do not involve using a stand-alone cooling module in comparison. In particular, we compared the radiative model against setups with a density-dependent $\beta(\Sigma)$ designed to quickly drop inside the cavity using a simple exponential cut-off, as well as a setup where $\beta$ is computed self-consistently through Eq.~\eqref{eq:beta_rad} by taking into account the local density, temperature, and opacity of each cell.

We found that both adaptive models are approximately 30\% faster compared to the radiative model, maintaining the same timestep but requiring significantly less computational time per step. The $\beta(\Sigma)$ model was marginally faster than the $\beta(\Sigma,T)$ model by 3--5\%, as it did not require the use of an opacity function, which case can be quite computationally expensive.

We note that the $\beta(\Sigma,T)$ model resulted in temperatures noticeably higher than in the radiative case. Nevertheless, a $\beta(\Sigma)$ model with $\beta_\text{ref}\rightarrow10^3$ instead of the used $10^2$ would result in a hotter disc as well (see Eq.~\eqref{eq:equilibrium-temperature}). Hence, a $\beta(\Sigma)$ model depends on an appropriate choice for $\beta_\text{ref}$ which introduces additional uncertainty. For a larger-scale system where irradiation dominates over viscous heating such that $\beta$ would be of the order of $10^{-1}$--$10^2$, the $\beta(\Sigma,T)$ model would approach the correct radiative profile more accurately while maintaining its significant speed-up factor over it.
Nevertheless, a $\beta(\Sigma)$ model would most likely also provide comparable results if a radius-dependent $\beta_\text{ref}$ is used. Of course, in both cases, a radiative model would provide the best results at the cost of a moderate computational overhead.

Based on our results, we stress that the choice of the target aspect ratio $h_0$ is important as it will dictate the aspect ratio in equilibrium $h_\text{eq}$ through Eq.~\eqref{eq:equilibrium-temperature}. It might be possible to reach a better agreement between $\beta$ and radiative models by intentionally choosing a lower $h_0$ such that $h_\text{eq}\approx0.044$. Nevertheless, our adaptive models match the radiative models in terms of cavity shape reasonably well even with our prescription of $h_0$.

\section{Conclusions}
\label{sec:conclusions}

In this study we have investigated three thermodynamical models in the context of circumbinary discs using 2D numerical simulations of locally isothermal, $\beta$-cooled, and radiative models.
The circumbinary environment naturally introduces a large inner cavity, the size and shape of which depends on the specific prescription of disc and binary parameters.

We find that the cavity itself develops to comparable sizes and shapes between the locally isothermal and radiative models. This happens because in the very low-density, optically thin cavity the cooling timescale of the radiative model is short enough to be comparable to a locally isothermal cooling prescription. However, the structure inside the inner disc and the width of the high-density region at the cavity edge differs between locally isothermal and radiatively cooled models. This divergence in the inner disc is also reflected in the different precession rates between the locally isothermal and radiative models.

By using $\beta$ models that simulate conditions of a nearly locally isothermal disc ($\beta\ll1$), we reproduce results that match the locally isothermal model well. However, when naively using a high value of $\beta=100$ to match the radiatively cooled prescription, models can deviate significantly. The reason is an unphysical treatment of the cooling timescale inside the cavity, leading to overheating of the inner disc. Such excess heat leads to an overly strong pressure increase in the inner disc, pushing the disc farther towards the binary. This leads to circularisation of the cavity especially for $e\bin=0$ as the wakes caused by the binary motion are no longer able to overcome the pressure-supported inner disc structure.

We introduce two possible solutions of adaptive $\beta$ prescriptions to achieve parametrised models that can potentially be more physically accurate than constant $\beta$ models: an empirical, opacity-model-agnostic prescription in which densities below a threshold value lead to a drop in the cooling timescale such that the effective $\beta(\Sigma)$ profile emulates both the optically thin cavity and optically thick disk as would be expected in a radiative model, and a more physically motivated model where $\beta(\Sigma,T)$ is evaluated locally in a self-consistent manner by computing the thermal cooling timescale based on the current density, temperature, and opacity of each cell. We found that the first model produces reasonably good results in terms of both temperature profiles and cavity properties, while the second model matches the cavity shape of our radiative runs more closely at the cost of a slightly hotter disc far from the binary. The latter happens due to the fundamentally different approach of a relaxation prescription using $\beta$---where the background temperature profile is assumed to be known in advance---as opposed to a truly radiative setup where the gas could, in principle, cool down to very low temperatures. The differences between our prescriptions of adaptive $\beta$ and radiative models underline the importance of correctly modelling the binary cavity when comparing to observations.

After investigating models with different, constant cooling timescales $\beta\in[10^{-2}, 10^2]$ we find that the size and eccentricity of the cavity becomes minimal for values of $\beta$ close to 1. This is likely due to the same effect observed in the work by \cite{2020Miranda_II}, where radiative cooling of the spiral arms launched by the binary interferes with the angular momentum transport capabilities of said spirals for $\beta\sim1$, thereby resulting in smaller and more circular cavities. We note that our models did not include the effects of in-plane cooling, which might affect our results according to their study.

\begin{acknowledgements}
We dedicate this paper to our late mentor and friend Willy Kley. Without his support and guidance, this work would not have been possible. We will always be grateful for all the time he invested in teaching us and helping us grow.
AP was funded by  grant 285676328 of the German Research Foundation (DFG).
The authors acknowledge support by the High Performance and
Cloud Computing Group at the Zentrum f\"ur Datenverarbeitung of the University
of T\"ubingen, the state of Baden-W\"urttemberg through bwHPC, and the German
Research Foundation (DFG) through grant no INST 37/935-1 FUGG. AZ and RPN are supported by STFC grant ST/P000592/1, and RPN is supported by the Leverhulme Trust through grant RPG-2018-418.
All plots in this paper were made using the Python library \texttt{matplotlib} \citep{Hunter:2007}.
\end{acknowledgements}

\newpage
\bibliography{beta-cbd}
\bibliographystyle{aa}

\newpage
\phantom{break page}
\newpage

\begin{appendix}
\section{Highly viscous environments}

\begin{figure}[h]
	\centering
	\includegraphics[width=\textwidth]{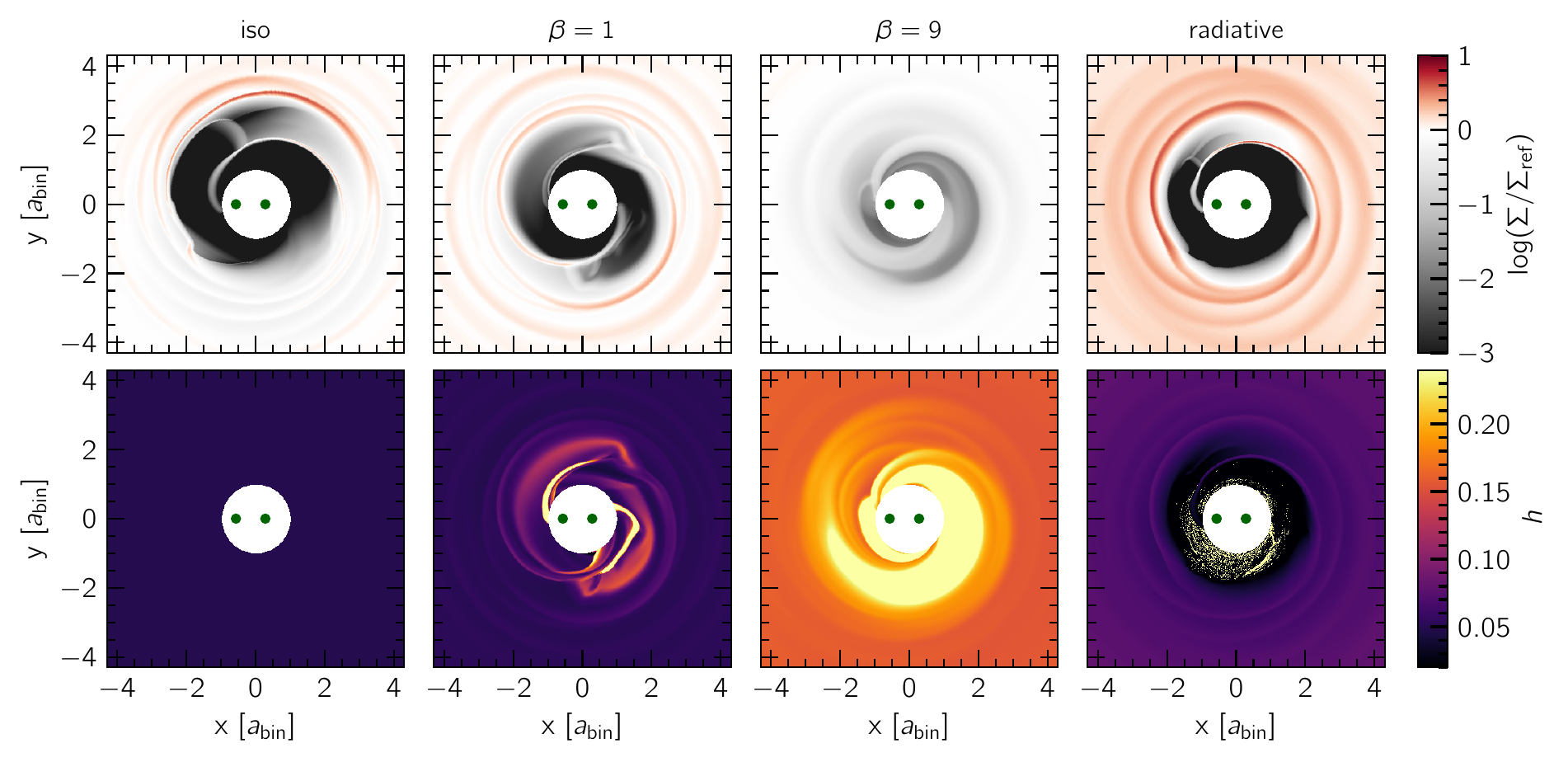}
	\parbox{\textwidth}{\caption{Comparisons of surface density (top) and aspect ratio (bottom) heatmaps for a viscous $\alpha=0.1$ between (from left to right) isothermal, $\beta=1$, $\beta=9$ and radiative models for $e\bin=0.15$, at time $700~T\bin$. The top panels have been normalised to their reference power-law profiles $\Sigma_\text{ref} \propto r^{-0.5}$.}}
	\label{fig:alpha-2d}
\end{figure}

Accretion discs around black holes are believed to be much more viscous than typical PPDs. 
For a first idea on the effect of cooling in such systems, we tested out a small series of models using the parameters given in Sec.~\ref{init} with an $e\bin=0.15$ and a high Shakura--Sunyaev parameter of $\alpha=0.1$. For this setup, we ran models with locally isothermal, radiative, and $\beta=[1,9]$ prescriptions. Owing to the shorter viscous timescale for this choice of $\alpha$, simulations already show a fully converged state at $t=700~T\bin$. Fig.~\ref{fig:alpha-2d} shows the density and thermal structure of the inner disc.

We see that for highly viscous disc models, the $\beta$ prescription is ill-suited to simulate the inner cavity structure.
There are minor differences, with the locally isothermal model opening the widest eccentric cavity. In contrast, the radiative model shows a slightly smaller and diffused cavity due to the heating induced by the additional viscosity.
However, the $\beta=1$ model shows a more shallow and almost completely symmetric cavity that does not compare well with the structure of the locally isothermal and radiative models. Moreover, the aspect ratio of the material and the streams inside the cavity increases to unrealistic levels. The $\beta=9$ case does not reproduce the locally isothermal or radiative model appropriately. Cooling in the inner region is slow by construction such that the cavity growth is almost completely inhibited. As a result, the overall temperatures in the disc rise to extreme levels (see Eq.~\eqref{eq:equilibrium-temperature}), puffing up  the disc to aspect ratios of nearly 0.2, which cannot describe the bulk of the disc realistically.

For such extremely viscous environments, treating radiative effects can produce different results depending on the exact physical parameters compared to the often-used locally isothermal prescription. However, a $\beta$-cooling model should not be used for such systems.

\vfill

%alt version: new page
%\begin{figure*}[h]
%	\centering
%	\includegraphics[width=\textwidth]{2d_alpha}
%	\caption{Comparisons of surface density (top) and aspect ratio (bottom) heatmaps for a viscous $\alpha=0.1$ between (from left to right) isothermal, $\beta=1$, $\beta=9$ and radiative models for $e\bin=0.15$, at time $700~T\bin$. The top panels have been normalised to their reference power-law profiles $\Sigma_\text{ref} \propto r^{-0.5}$.}
%	\label{fig:alpha-2d}
%\end{figure*}

\end{appendix}
\end{document}